\documentclass[aps,prd,twocolumn,superscriptaddress]{revtex4-2}

\usepackage{amsmath}
\usepackage{tikz}
\usepackage{multirow}

\begin{document}


\title{Contact Interactions at Future Circular Collider based Muon-Proton Colliders}


\author{Gural Aydin}
\email[]{gaydin@mku.edu.tr}
\affiliation{ Department of Physics, Hatay Mustafa Kemal University, Hatay, Turkey}

\author{Yusuf O. Günaydin}
\email{yusufgunaydin@gmail.com}

\affiliation{Department of Physics, Kahramanmaras Sütcü Imam University, Kahramanmaras, Turkey}

\author{Mehmet Sahin}
\email{mehmet.sahin@usak.edu.tr}

\affiliation{Department of Computer Engineering, Usak University, Usak, Turkey}

\author{Saleh Sultansoy}
\email{ssultansoy@etu.edu.tr}

\affiliation{TOBB Economics and Technology University, Ankara, Turkey}

\affiliation{ANAS, Institute of Physics, Baku, Azerbajian}
\author{Mehmet T. Tarakcioglu}
\email[]{turkertarakcioglu@gmail.com}
\affiliation{Department of Physics, Kahramanmaras Sütcü Imam University, Kahramanmaras, Turkey}



\begin{abstract}
Recently proposed Future Circular Collider based muon-proton colliders will allow investigating lepton-hadron interactions at the highest center-of-mass energy. In this study, we investigate the potential of these colliders for a four-fermion contact interactions search. Regarding the constructive and destructive interferences of contact interactions, we estimated discovery, observation, and exclusion limits on the compositeness scale for the left-left, right-right, left-right, and right-left helicity structures. This study’s findings show that the FCC-based $\mu p$ colliders have great potential for investigating four-fermion contact interactions.	
	
\end{abstract}


\maketitle

\section{\label{sec:int}Introduction}

The Standard Model (SM) is a theory that remarkably describes elementary particles and their strong, weak, and electromagnetic interactions. It also shows a good agreement with the experimental results on High Energy Physics. Thus, the Standard Model gives answers to many questions about our universe. However, there are some questions that SM cannot answer. For example, why quarks and leptons repeat in families, why the Standard Model has so many parameters, how neutrinos gain their masses, and so on.  To answer these questions, new theories beyond Standard Model have emerged. Among these theories, Composite Models \cite{pati1974,terazawa1977,harari1979,shupe1979,terazawa1980,fritzsch1981,terazawa1982,eichten1983,lyons1983,terazawa1983,dsouza1992,celikel1998,desouza2008,terazawa2014,fritzsch2016,terazawa2015,Kaya2018} can respond to the best pattern to reduce the Standard Model’s parameter redundancy.

If SM fermions have a composite structure, they may consist of more fundamental particles called preons. The new physics scale on which the preons will emerge is called the compositeness scale  ($\Lambda$). Suppose the particle colliders’ subprocess energy is greater than the compositeness scale of fermions. In that case, research on compositeness can be done directly at particle colliders. On the other hand, research on compositeness can be performed indirectly through contact interactions if the colliders’ subprocess energy is smaller than the compositeness scale. In literature, there are some studies on contact interactions \cite{eichten1983,eichten1984,Buchmuller:1987ur,Schrempp:1987zy,Argyres:1990zs,Rizzo:1994sk,barger1998,cheung1998muon,zarnecki1999,Babich:2001ik,pgd2020:aa}.   

Contact interactions investigations were performed at electron-positron  \cite{aleph2006:aa,delphi2006:aa,delphi2009:aa,acciarri2000,opal2004:aa}, electron-proton \cite{aaron2011:aa,chekanov2004}, and hadron colliders \cite{cdf1997:aa,
	cdf2001:aa,
	cdf2006:aa,
	dZero1999:aa,
	dZero2009:aa,
	atlas2011:aa,
	atlas2011:ab,
	atlas2013:ab,
	Aad_2014,
	atlas2015:aa,
	atlas2016:aa,
	atlas2017:aa,
	cms2010:aa,
	cms2011:aa,
	cms2012:aa,
	cms2015:aa,
	cms2013:aa,
	cms2017:aa,
	cms2018:aa} experiments. If SM leptons and quarks are composite structures, $llqq$-type four fermion contact interactions occur.  Here, $l$ and $q$  represent electron/muon and quarks, respectively. Using 36 fb\textsuperscript{-1} data-set at $\sqrt{s}$ = 13 TeV, the ATLAS  Collaboration puts exclusion limits on the $llqq$-type contact interaction scale in the $qq \rightarrow ll$ process \cite{atlas2017:ab}. Contact interaction scales of the $llqq$-type for constructive (destructive) interference are excluded as $\Lambda$ = 35 (28) TeV and below for the right-right helicity structure, $\Lambda$ = 40 (25) TeV and below for the left-left helicity structure. The ATLAS Collaboration also  excluded the $llqq$-type contact interaction scales $\Lambda$ = 36 (28) TeV and below for the left-right helicity structure with constructive (destructive) interference. Ditto, the CMS Collaboration puts exclusion limits on the compositeness scale  as 20 TeV and 32 TeV for left-left destructive  and right-right constructive cases, respectively \cite{cms2018:aa}. 

In this paper, we investigated contact interactions at Future Circular Collider (FCC) based muon-proton colliders. In section II, we give the main parameters of the FCC-based muon-proton colliders. The following section presents Lagrangian of the contact interactions. Section IV includes transverse momentum and pseudo-rapidity distributions that determine applied cuts in our calculations.  Discovery, observation, and exclusion limit results for the compositeness scale are presented in section V.  Our conclusion is given in the last section.

\section{\label{sec:fcclh} The FCC-Based Muon-Proton Colliders}
The Future Circular Collider, built after the Large Hadron Collider has completed its runtime, is considered an energy-frontier machine by the high energy physics community. Besides proton-proton collisions, electron-proton and electron-positron collision experiments are also envisaged in the FCC \cite{fcc2019fcc,fcc2019fcc:ee,fcc2019fcc:hh,fcc2019he}.  
Furthermore, new solutions to the technical problems faced by muon colliders have attracted the attention of physicists upon the muon-proton colliders again \cite{neuffer_principles_1983,antonelli_novel_2016,zimmermann_lhc/fcc-based_2018,neuffer_feasibility_2018,boscolo2019future,delahaye2019muon,bartosik_muon_2019,mice_collaboration_demonstration_2020,ryne:muon2020,amapane2020,Long_2021}. Some advantages of muon-proton colliders over other colliders can be mentioned as the reason for this orientation. First, the synchrotron radiation problem, which is encountered at very high beam energies in electron-proton colliders, is eliminated in muon-proton colliders because the muon has a heavy mass relative to the electron. Therefore, at the multi-TeV center-of-mass energy level, muon-proton colliders can be advantageous for producing new TeV-scale particles in the mass shell. Moreover, muon-proton colliders may have a lower QCD background than proton-proton colliders in the BSM studies \cite{cheung2021}. Thus, contact interactions can be investigated more precisely at the multi-TeV scale in muon-proton colliders. Construction of the muon collider (or dedicated $\mu$-ring) tangential to FCC, as proposed in \cite{Acar_2018}, will allow handling the highest center-of-mass energy lepton-hadron collider.

Table  \ref{tab:fccMup} presents the main parameters of the FCC-based muon-proton colliders for four different muon beam energies. In the FCC, colliding proton beam energy will be 50 TeV. 

\begin{table}[ph]
	\caption{Basic parameters of the FCC-based $\mu p $ colliders \cite{acar2017}}
	\begin{center}
		{\begin{tabular}{@{}lccc@{}}\hline
				Collider Name &  E$_{\mu}$ (TeV)  & $\sqrt{s}$ (TeV)	& $\mathcal{L}_{int}$ (fb\textsuperscript{-1}/year ) \\ \hline
				$\mu$750$\otimes$FCC & 0.75  & 12.2 & 5 \\
				$\mu$1500$\otimes$FCC  & 1.50 & 17.3 & 5 \\
				$\mu$3000$\otimes$FCC  & 3.00  & 24.5 & 5 \\
				$\mu$20000$\otimes$FCC  & 20.0 & 63.2 & 10 \\ \hline
			\end{tabular}\label{tab:fccMup}}
	\end{center}
\end{table}

FCC-based $\mu p$ collider has been expected to run for ten years.  At the end of this 10-year run time, the $\mu$750$\otimes$FCC, $\mu$1500$\otimes$FCC, and $\mu$3000$\otimes$FCC  colliders will reach an integrated luminosity of 50 fb\textsuperscript{-1}, and the $\mu$20000$\otimes$FC collider of 100 fb\textsuperscript{-1}.   

Recently, the physics potential of the FCC-based $\mu p$ colliders has been investigated in many papers \cite{Caliskan_2017,TAKEUCHI2017, acar2018,Alici_2019,Ozansoy2019,Spor_2020,cheung2021}.

\section{\label{sec:intlag} Contact  Interaction Lagrangian}
If fermions have a substructure, they should have a new type of interaction. Investigating these interactions depends on the center-of-mass energy of the colliders and the compositeness scale. If the compositeness scale is much greater than the center-of-mass energy of the collider, the best method to investigate these phenomena would be through four fermion contact interactions. These interactions' most general flavor-diagonal chirally invariant Lagrangian \cite{eichten1983,eichten1984,pgd2020:aa} is described as

$$\mathcal{L} =  \mathcal{L}_{L L}  + \mathcal{L}_{R R} +\mathcal{L}_{L R} +\mathcal{L}_{R L}$$
with 
\begin{equation}
\label{eq:01}
\begin{aligned}
\mathcal{L}_{L L} &=\frac{g_{\text {contact }}^{2}}{2 \Lambda^{2}} \sum_{i, j} \eta_{L L}^{i j}\left(\bar{\psi}_{L}^{i} \gamma_{\mu} \psi_{L}^{i}\right)\left(\bar{\psi}_{L}^{j} \gamma^{\mu} \psi_{L}^{j}\right) \\
\mathcal{L}_{R R} &=\frac{g_{\text {contact }}^{2}}{2 \Lambda^{2}} \sum_{i, j} \eta_{R R}^{i j}\left(\bar{\psi}_{R}^{i} \gamma_{\mu} \psi_{R}^{i}\right)\left(\bar{\psi}_{R}^{j} \gamma^{\mu} \psi_{R}^{j}\right) \\
\mathcal{L}_{L R} &=\frac{g_{\text {contact }}^{2}}{2 \Lambda^{2}} \sum_{i, j} \eta_{L R}^{i j}\left(\bar{\psi}_{L}^{i} \gamma_{\mu} \psi_{L}^{i}\right)\left(\bar{\psi}_{R}^{j} \gamma^{\mu} \psi_{R}^{j}\right) \\
\mathcal{L}_{R L} &=\frac{g_{\text {contact }}^{2}}{2 \Lambda^{2}} \sum_{i, j} \eta_{R L}^{i j}\left(\bar{\psi}_{R}^{i} \gamma_{\mu} \psi_{R}^{i}\right)\left(\bar{\psi}_{L}^{j} \gamma^{\mu} \psi_{L}^{j}\right)
\end{aligned}
\end{equation}
where $g_{\text {contact }}^{2}$ is coupling constant ($g_{\text {contact }}^{2} = 4\pi$), $\Lambda$ is compositeness scale, $\eta_{L L}^{i j},\;\eta_{RR}^{i j},\; \eta_{LR}^{i j},$  and  $\eta_{RL}^{i j}$ are chirality coefficients, $\psi_{L}^{i},\; \psi_{L}^{j},\; \psi_{R}^{i}, \text{and} \;\psi_{R}^{j}$  are fermion spinors, $i$, $j$ represent the indices of fermion species, and $L$ and $R$ stand for left and right helicity, respectively. 

In this investigation, we regard four-fermion interactions ($\mu p \rightarrow \mu + j + X$) whose cross-section is described by $\sigma_{tot} = \sigma_{SM} - \eta_{ij} \dfrac{F_I}{\Lambda^2} + \dfrac{F_C}{\Lambda^4}$ \cite{Aad_2014}. The first term in this equation shows to the SM interactions, the second term relates to interference of the SM and four-fermion interactions, and the third term involves the contribution from pure contact interactions as a new physics (NP) only. Here, the parameters $F_I$ and $F_C$ are functions of the cross-section not dependent on $\Lambda$. As the compositeness scale value rises, the term standing for the interference of contact interactions with the SM comes dominant. Then, the leading term in this research is the term denoting four-fermion contact interactions with the SM.

We first implemented this Lagrangian into the CalcHEP \cite{calchep2013} simulation software via LanHEP \cite{LanHEP,lanhep2016}. Then, in numerical calculations, we used the following notations: 

\begin{equation}
\label{eq:02}
\begin{aligned}
\Lambda &= \Lambda_{LL}^\pm\; \text{for}\; (\eta_{L L}^{i j}, \eta_{RR}^{i j},\eta_{LR}^{i j},\eta_{RL}^{i j}) = (\pm1, 0,0,0)\\
\Lambda &= \Lambda_{RR}^\pm\; \text{for}\; (\eta_{L L}^{i j}, \eta_{RR}^{i j},\eta_{LR}^{i j},\eta_{RL}^{i j}) = (0,\pm1,0,0)\\
\Lambda &= \Lambda_{LR}^\pm\; \text{for}\; (\eta_{L L}^{i j}, \eta_{RR}^{i j},\eta_{LR}^{i j},\eta_{RL}^{i j}) = (0,0,\pm1,0)\\
\Lambda &= \Lambda_{RL}^\pm\; \text{for}\; (\eta_{L L}^{i j}, \eta_{RR}^{i j},\eta_{LR}^{i j},\eta_{RL}^{i j}) = (0,0,0,\pm1).
\end{aligned}
\end{equation}
\section{\label{sec:distr} Transverse Momentum and Pseudorapidty Distributions}

In this study, we investigated four fermion contact interactions at muon-proton colliders with different center-of-mass energies. We used CalcHEP simulation software in our calculations. We chose  $\mu$ + $p$ $\rightarrow$ $\mu$ + $j$ + X as a signal and as a background process. Difference between signal and background is that background does not have contact interaction vertices.  Here, $j$ denotes $ u, \bar{u}, d, \bar{d}, c, \bar{c}, s, \bar{s}, b, \; \text{and}\; \bar{b}$.  
Since the detection efficiency of jets with $p_T$ above 20 GeV is almost 100\%, any uncertainty originating from the selection of jets \cite{atlas2017:aa} will not be affected in our calculations with the cuts we applied. Muon beam decays in the collider ring that creates addition to background, which is called beam-induced-background (BIB). According to references \cite{Bartosik:2020xwr,Lu:2020dkx,Collamati:2021sbv}, BIB does not affect muon collider physics performance by some regulation of detectors.  Furthermore, the LHeC Collaboration reported systematic uncertainties originated from $\alpha_s$ and PDF are smaller than statistical uncertainties in the latest publication \cite{agostini2021}. So, we neglected systematic errors in our calculations because of statistical uncertainties domination over the systematics. For the quark distribution functions, we selected the CT10 \cite{lai:2010aa}, and for detector acceptance, we put  $P_{T_{jet,\mu}} >$ 25 GeV cuts on the transverse momentum of the muon and jet. In order to show the difference of the signal from the background, we obtained the transverse momentum and pseudorapidity plots by considering the CI+SM interactions as the signal process and only the SM interactions for the background.

\begin{figure}[h!]
	\resizebox{0.5\textwidth}{!}{\input{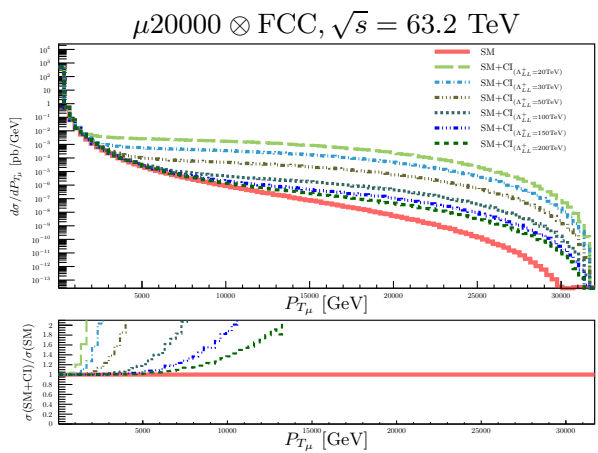}}
	\caption{\label{fig:transversemuonLL} Transverse momentum distribution for final state muon at 	$\mu$20000$\otimes$FCC  collider option with $\Lambda^+_{LL}$}
\end{figure}

\begin{figure}[h!]
	\resizebox{0.5\textwidth}{!}{\input{eTaplot_FCC_mup_nj_E20_LL.tex}}
	\caption{\label{fig:etajetLL} Pseudorapidity distribution for jet final state  at	$\mu$20000$\otimes$FCC  collider option with $\Lambda^+_{LL}$}
\end{figure}

\begin{table}[h!]
	\caption{\label{tab:cuts}Applied cut-sets according to distribution plots for each collider option}
	\begin{ruledtabular}
		\resizebox{0.5\textwidth}{!}{
		\begin{tabular}{c|c|l|c|l}
			\multirow{2}{*}{$\Lambda$} & \multicolumn{2}{c|}{\begin{tabular}[c]{@{}c@{}}	$\mu$750$\otimes$FCC \\ ($\sqrt{s}$ = 12.2 TeV)\end{tabular}}   & \multicolumn{2}{c}{\begin{tabular}[c]{@{}c@{}} $\mu$1500$\otimes$FCC \\($\sqrt{s}$ = 17.3 TeV)\end{tabular}}   \\ \cline{2-5} 
			& \multicolumn{1}{c|}{\begin{tabular}[c]{@{}c@{}}$P_{T_{jet,\mu}}$ (GeV)\end{tabular}} & \multicolumn{1}{c|}{$\eta_{jet}$} & \multicolumn{1}{c|}{\begin{tabular}[c]{@{}c@{}}$P_{T_{jet,\mu}}$ (GeV)\end{tabular}}    &\multicolumn{1}{c}{$\eta_{jet}$}                  \\ \hline
			$\Lambda^+_{LL}$           & $>$2000        & -4.5$<\eta<$1         & $>$3500     & -4.5$<\eta<$1      \\
			$\Lambda^-_{LL}$           & $>$4000        & -4.5 $<\eta<$0.5     & $>$5500     & -4.5$<\eta<$1       \\
			$\Lambda^+_{RR}$           & $>$2000       & -4.5 $<\eta<$1        & $>$3500     & -4.5$<\eta<$1       \\
			$\Lambda^-_{RR}$           & $>$4000       & -4.0 $<\eta<$1        & $>$5500     & -4.0$<\eta<$1         \\
			$\Lambda^+_{LR}$           & $>$1500       & -4.5 $<\eta<$1        & $>$2500     & -5.0$<\eta<$1         \\
			$\Lambda^-_{LR}$           & $>$3000       & -4.0 $<\eta<$1        & $>$4000      & -4.0$<\eta<$1         \\
			$\Lambda^+_{RL}$           & $>$1750      & -4.5 $<\eta<$1        & $>$3000      & -4.5$<\eta<$1          \\
			$\Lambda^-_{RL}$           & $>$3000       & -4.5 $<\eta<$1        & $>$4000      & -4.5$<\eta<$1          \\ \hline\hline
			\multirow{2}{*}{$\Lambda$} & \multicolumn{2}{c|}{\begin{tabular}[c]{@{}c@{}}	$\mu$3000$\otimes$FCC \\ ($\sqrt{s}$ = 24.5 TeV)\end{tabular}}   & \multicolumn{2}{c}{\begin{tabular}[c]{@{}c@{}} $\mu$20000$\otimes$FCC \\($\sqrt{s}$ = 63.2 TeV)\end{tabular}}   \\ \cline{2-5} 
			& \multicolumn{1}{c|}{\begin{tabular}[c]{@{}c@{}}$P_{T_{jet,\mu}}$ (GeV)\end{tabular}}  & \multicolumn{1}{c|}{$\eta_{jet}$}  & \multicolumn{1}{c|}{\begin{tabular}[c]{@{}c@{}}$P_{T_{jet,\mu}}$ (GeV)\end{tabular}} & \multicolumn{1}{c}{$\eta_{jet}$} \\ \hline
			$\Lambda^+_{LL}$           & $>$4000     & -4.5$<\eta<$1    & $>$12000   & -4.5$<\eta<$2      \\
			$\Lambda^-_{LL}$           & $>$6000     & -4.5$<\eta<$1    & $>$16000    & -5.0$<\eta<$2      \\
			$\Lambda^+_{RR}$           & $>$3500    & -4.5$<\eta<$1    & $>$11000    & -5.0$<\eta<$2      \\
			$\Lambda^-_{RR}$           & $>$6000     & -4.0$<\eta<$1    & $>$16000   & -4.5$<\eta<$1       \\
			$\Lambda^+_{LR}$           & $>$2500    & -5.0$<\eta<$1    & $>$8000     & -5.0$<\eta<$2        \\
			$\Lambda^-_{LR}$           & $>$3500    & -4.5$<\eta<$1     & $>$13000   & -4.5$<\eta<$1       \\
			$\Lambda^+_{RL}$           & $>$3000    & -4.5$<\eta<$1     & $>$10000   & -5.0$<\eta<$1       \\
			$\Lambda^-_{RL}$           & $>$3500     & -4.5$<\eta<$1     & $>$12000    & -4.5$<\eta<$1          
		\end{tabular}
	}
	\end{ruledtabular}
\end{table}

\begin{figure}[h!]
	\resizebox{0.50\textwidth}{!}{\input{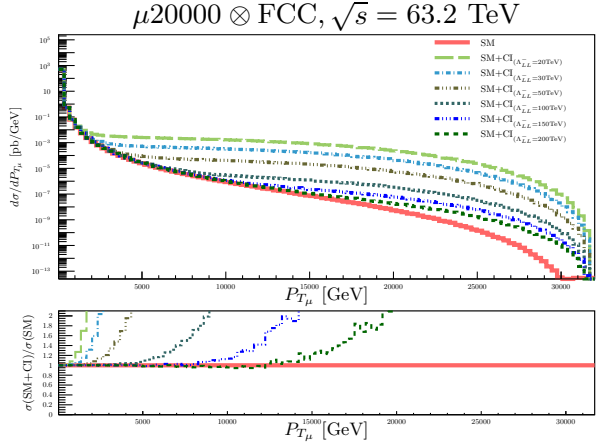}}
	\caption{\label{fig:transversemuonminusLL} Transverse momentum distribution for final state muon at $\mu$20000$\otimes$FCC collider option with $\Lambda^-_{LL}$}
\end{figure}

Among these distribution plots, we have presented the transverse momentum and pseudorapidity distributions for the collider with 63.2 TeV center-of-mass energy, one of the four collider options, considering constructive $LL$ and destructive $LL$ interferences as an example. As seen in Figure \ref{fig:transversemuonLL}, the point at which the $\sigma_{(CI+SM)}$ distribution begins to separate clearly from the $\sigma_{(SM)}$ distribution is 12000 GeV. Thus, we set the cut limit $P_{T_\mu} >$ 12000 GeV in constructive interference for the transverse momentum of the final state muon. Since the transverse momentum distribution function of the final state jet also behaves exactly like the muon, we set the same cut limit for the jet. Figure \ref{fig:etajetLL} shows the pseudorapidity distribution in the constructive interference for the final state jet. According to this graph, we determined the cut  on pseudorapidity  as -4.5 $ <\eta_ {jet}<$ 2 intervals  via checking the region where the CI+SM and SM distributions differ from each other. Since the CI+SM and SM pseudorapidity distributions of the final state muon show the same pattern, we chose the range -2.5 $ <\eta_\mu<$ 2.5 for  the pseudorapidity cut limit. Using Figures \ref{fig:transversemuonminusLL} and \ref{fig:etajetminusLL}, for destructive interferences, we similarly set cut limits as  $P_{T_{\mu}}>$ 16000 GeV, $P_{T_{jet}}>$ 16000 GeV, -5 $<\eta_{jet}<$ 2,  and -2.5$<\eta_\mu<$ 2.5. Likewise, we checked transverse momentum and pseudorapidity distributions for the rest of the chiralities with constructive and destructive interferences. Similar analyses have been performed for  each collider options, and specified cut limits are listed in Table \ref{tab:cuts}.

\begin{figure}[h!]
	\resizebox{0.50\textwidth}{!}{\input{eTaplot_FCC_mup_nj_minus_E20_LL.tex}}
	\caption{\label{fig:etajetminusLL} Pseudorapidity distribution for jet final state  at 	$\mu$20000$\otimes$FCC collider option with $\Lambda^-_{LL}$}
\end{figure}

\begin{table}[h!]
		\caption{\label{tab:events}Cut-flow table for $\sqrt{s}$=12.2 TeV option with -2.5 $< \eta_\mu <$ 2.5.}
			\begin{ruledtabular}
		\resizebox{0.5\textwidth}{!}{
	\begin{tabular}{lccc|ccc}
		   & \multicolumn{6}{c}{Number of Events}                                  \\ \cline{2-7}
	 & \multicolumn{3}{c|}{SM}            & \multicolumn{3}{c}{CI+SM}         \\ \cline{2-7}
	 Cuts:	& No Cut & $\eta_{jet,\mu}$  & $P_{T_{jet,\mu}}$ & No Cut & $\eta_{jet,\mu}$   & $P_{T_{jet,\mu}}$ \\ \cline{2-7}
		$LL+$  & 7.52$\times$10$^{8}$   & 6215    & 467        & 7.52$\times$10$^{8}$   & 7685    & 1106       \\
		$LL-$                       & 7.52$\times$10$^{8}$   & 6215    & 4          & 7.52$\times$10$^{8}$   & 5950    & 27         \\
		$RR+$                       & 7.52$\times$10$^{8}$   & 6215    & 467        & 7.52$\times$10$^{8}$   & 7655    & 1062       \\
		$RR-$                       & 7.52$\times$10$^{8}$   & 1179    & 4          & 7.52$\times$10$^{8}$   & 1240    & 30         \\
		$LR+$                       & 7.52$\times$10$^{8}$   & 6215    & 1594       & 7.52$\times$10$^{8}$   & 7025    & 2075       \\
		$LR-$                       & 7.52$\times$10$^{8}$   & 1179    & 48         & 7.52$\times$10$^{8}$   & 1333    & 106        \\
		$RL+$                       & 7.52$\times$10$^{8}$   & 6215    & 847        & 7.52$\times$10$^{8}$   & 6965    & 1219       \\
		$RL-$                       & 7.52$\times$10$^{8}$   & 6215    & 48         & 7.52$\times$10$^{8}$   & 6325    & 104     \\ \hline
		\multicolumn{7}{l}{$\Lambda$ = 30 TeV; $\mathcal{L}_{int}$ = 50 fb$^{-1}$} 
	\end{tabular}
}
	\end{ruledtabular}
\end{table}

To show applied cuts' effects on the number of events on both the CI+SM (signal) and the SM (background), we included the cut-flow table for the $\mu$750$\otimes$FCC collider option as an example. Table \ref{tab:events} illustrates the impact of cuts from Table \ref{tab:cuts} on the number of events for the  $\mu$750$\otimes$FCC collider option. It is apparent from Table \ref{tab:events}  that after applying all cuts, the signal  events become more distinguishable than the background events. 

\section{Significance Calculation for Compositeness Scale}
In this section, the calculation results for the  exclusion (2$\sigma$), observation (3$\sigma$), and discovery (5$\sigma$) limits on the compositeness scale in contact interactions at FCC-based muon-proton colliders are given. For this, we used equation \ref{eq:signifi} to obtain statistical significances calculation for both constructive and destructive interferences. 

\begin{equation}
\label{eq:signifi}
Significance = \dfrac{\sigma_{(\text{CI+SM})}-\sigma_{(\text{SM})}}{\sqrt{\sigma_{(\text{SM})}}}\sqrt{\mathcal{L}_{int}}
\end{equation}

Here, $\sigma_{(\text{CI+SM})}$ denotes Contact and Standard Model interactions cross section as a signal,  $\sigma_{(\text{SM})}$ represents Standard Model cross section as a background,  and $\mathcal{L}_{int}$ is integrated luminosity of the $\mu p$ colliders. The statistical uncertainties in our calculations were because of uncertainty in the integrated luminosity. $\delta \mathcal{L} / \mathcal{L} = 1/\sqrt{\mathcal{L}}$ equality scales the sensitivity of the integrated luminosity \cite{Abramowicz:1443828,Cepeda:2019klc}. Therefore, using this sensitivity in our significance calculations, we calculated the statistical uncertainties in the compositeness scale.

\begin{center}
	
	\begin{figure}[h!]
		\resizebox{0.50\textwidth}{!}{\input{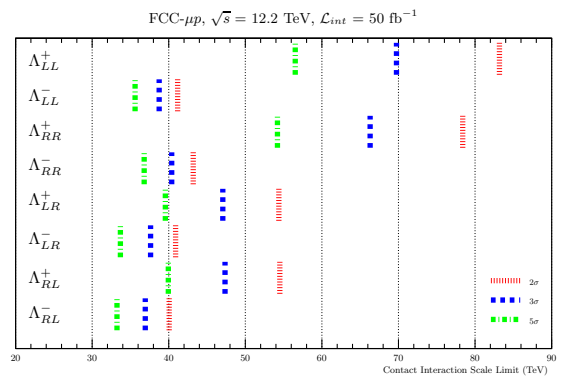}}	
		\caption{Contact interactions scale limits for FCC-based muon-proton collider with $\sqrt{s} $ = 12.2 TeV at $\mathcal{L}_{int}$ = 50 fb\textsuperscript{-1}.\label{fig:ci_mup750}}
	\end{figure}
\end{center}

\onecolumngrid


\begin{table}[]
	\caption{Attainable limits on the compositeness scales with sensitivity at the FCC-based $\mu p$ collider with $\sqrt{s}$ = 12.2 TeV for the first, the  fifth, and the tenth years' luminosities. }
	\resizebox{\textwidth}{!}{\begin{tabular}{@{}cccccccccc@{}} \hline
			\multicolumn{2}{l}{\textbf{$\mu$750$\otimes$FCC}} & 	\multicolumn{8}{c}{$\Lambda \pm \delta \%$ (TeV)}\\ \hline
			\begin{tabular}[c]{@{}c@{}}$\mathcal{L}_{int}$ \\ (fb\textsuperscript{-1})\end{tabular}& CL & $\Lambda^{+}_{LL}$ & $\Lambda^{-}_{LL}$ & $\Lambda^{+}_{RR}$& $\Lambda^{-}_{RR}$ & $\Lambda^{+}_{LR}$ & $\Lambda^{-}_{LR}$ & $\Lambda^{+}_{RL}$ & $\Lambda^{-}_{RL}$ \\ \hline
			\multirow{3}{*}{5} &
			5$\sigma$ &
			37.0 ± 4.2\% &
			28.4 ± 2.8\% &
			35.5 ± 4.4\% &
			29.0 ± 2.7\% &
			27.7 ± 3.7\% &
			25.9 ± 3.0\% &
			28.2 ± 3.6\% &
			25.7 ± 2.7\% \\
			&
			3$\sigma$ &
			44.3 ± 4.5\% &
			31.6 ± 2.5\% &
			42.6 ± 4.5\% &
			32.4 ± 2.6\% &
			32.3 ± 3.8\% &
			29.2 ± 2.8\% &
			32.8 ± 3.8\% &
			28.9 ± 2.8\% \\
			&
			2$\sigma$ &
			51.5 ± 4.7\% &
			34.1 ± 2.3\% &
			49.5 ± 4.6\% &
			35.2 ± 2.5\% &
			36.7 ± 3.9\% &
			32.0 ± 2.8\% &
			37.1 ± 3.9\% &
			31.6 ± 2.8\% \\ \hline
			\multirow{3}{*}{25} &
			5$\sigma$ &
			49.4 ± 3.1\% &
			33.5 ± 1.5\% &
			47.5 ± 3.0\% &
			34.4 ± 1.7\% &
			35.4 ± 2.6\% &
			31.2 ± 1.9\% &
			35.8 ± 2.4\% &
			30.9 ± 1.8\% \\
			&
			3$\sigma$ &
			60.4 ± 3.3\% &
			36.6 ± 1.4\% &
			57.8 ± 3.2\% &
			37.9 ± 1.5\% &
			41.8 ± 2.7\% &
			34.9 ± 1.8\% &
			42.1 ± 2.7\% &
			34.4 ± 1.7\% \\
			&
			2$\sigma$ &
			71.5 ± 3.5\% &
			39.1 ± 1.3\% &
			67.9 ± 3.3\% &
			40.8 ± 1.4\% &
			48.0 ± 2.9\% &
			38.1 ± 1.7\% &
			48.3 ± 2.8\% &
			37.4 ± 1.7\% \\ \hline
			\multirow{3}{*}{50} &
			5$\sigma$ &
			56.5 ± 2.7\% &
			35.6 ± 1.2\% &
			54.2 ± 2.6\% &
			36.8 ± 1.3\% &
			39.6 ± 2.2\% &
			33.7 ± 1.5\% &
			39.9 ± 2.2\% &
			33.2 ± 1.5\% \\
			&
			3$\sigma$ &
			69.7 ± 2.9\% &
			38.7 ± 1.1\% &
			66.3 ± 2.8\% &
			40.4 ± 1.2\% &
			47.0 ± 2.4\% &
			37.6 ± 1.4\% &
			47.4 ± 2.3\% &
			36.9 ± 1.4\% \\
			&
			2$\sigma$ &
			83.2 ± 3.0\% &
			41.2 ± 1.0\% &
			78.4 ± 2.9\% &
			43.2 ± 1.1\% &
			54.4 ± 2.5\% &
			40.9 ± 1.4\% &
			54.5 ± 2.4\% &
			40.1 ± 1.4\%\\ \hline
		\end{tabular}\label{tab:fccmu750}}
\end{table}



\twocolumngrid

\begin{center}
	\begin{figure}[h!]
		\resizebox{0.5\textwidth}{!}{\input{ci_mup1500.tex}}	
		\caption{Contact interactions scale limits for FCC-based muon-proton collider with $\sqrt{s} $ = 17.3 TeV at $\mathcal{L}_{int}$ = 50 fb$^{-1}$.\label{fig:ci_mup1500}}
	\end{figure}
\end{center}

\begin{center}
	\begin{figure}[h!]
		\resizebox{0.5\textwidth}{!}{\input{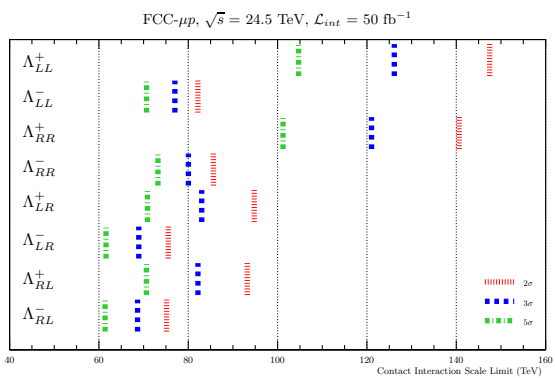}}	
		\caption{Contact interactions scale limits for FCC-based muon-proton collider with $\sqrt{s} $ = 24.5 TeV at $\mathcal{L}_{int}$ = 50 fb$^{-1}$.\label{fig:ci_mup3000}}
	\end{figure}	
\end{center}

Using the relevant cut sets in Table \ref{tab:cuts} and equation \ref{eq:signifi}, the exclusion, observation, and discovery limits of compositeness scale for all constructive and destructive interferences ($\Lambda^{+}_{LL}$, $\Lambda^{-}_{LL}$, $\Lambda^{+}_{RR}$, $\Lambda^{-}_{RR}$, $\Lambda^{+}_{LR}$, $\Lambda^{-}_{LR}$, $\Lambda^{+}_{RL}$, and $\Lambda^{-}_{RL}$ ) for the final luminosity value of 50 fb\textsuperscript{-1} at the muon-proton collider with center-of-mass energy $\sqrt{s}$ = 12.2 TeV are given in figure \ref{fig:ci_mup750}. It is seen that  $\Lambda^{+}_{LL}$ has the highest compositeness scale limits, but $\Lambda^{-}_{RL}$ has the lowest.  The compositeness scale values with sensitivity lie between 33.2 $\pm$ 1.5\% TeV and 56.5 $\pm$ 2.7\% TeV for the discovery, 36.9 $\pm$ 1.4\% TeV and 69.7 $\pm$ 2.9\% TeV for observation, and 40.1 $\pm$ 1.4\% TeV and 83.2 $\pm$ 3.0\%  TeV for exclusion. These limits are far beyond the LHC experimental results. Compositeness scale limits in terms of luminosities are given in Table  \ref{tab:fccmu750} for the rest of the helicities with constructive and destructive interferences.

\begin{center}
	\begin{figure}[h!]
		\resizebox{0.5\textwidth}{!}{\input{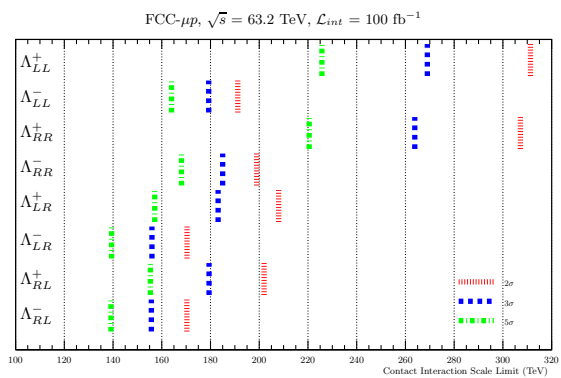}}
		\caption{Contact interactions scale limits for FCC-based muon-proton collider with $\sqrt{s}$ = 63.2 TeV at $\mathcal{L}_{int}$ = 100 fb$^{-1}$.\label{fig:ci_mup20000}}
	\end{figure}	
\end{center}
As mentioned above, we used the relevant cut sets in Table \ref{tab:cuts} and equation 3 to calculate statistical significances for the other two collider options with 17.3 and 24.5 TeV center-of-mass energies. Figure \ref{fig:ci_mup1500} and \ref{fig:ci_mup3000} shows the compositeness scale limits for all constructive and destructive interferences, respectively. The highest limit is achieved for  $\Lambda^{+}_{LL}$, and the lowest limit is achieved  for $\Lambda^{-}_{RL}$  in both collider options. As seen in figure \ref{fig:ci_mup1500}, discovery limits show variation between 45.9 $\pm$ 1.4\%  TeV and 78.4 $\pm$ 2.5\%   TeV; observation limits are between 51.0 $\pm$ 1.4\%   TeV and 95.3 $\pm$ 2.7\% TeV; exclusion limits swing from 55.4 $\pm$ 1.4\% TeV to 112.3 $\pm$ 2.8\% TeV. In table \ref{tab:fccmu3000}, the highest and the lowest discovery, observation, and exclusion limits are (104.7 $\pm$ 2.4\%\,:\,61.4 $\pm$ 1.5\%) TeV, (126.1 $\pm$ 2.6\%\,:\,68.6 $\pm$ 1.5\%) TeV, and (147.5 $\pm$ 2.7\% \,:\,75.2 $\pm$ 1.5\%) TeV for the $\mu$3000$\otimes$FCC  option with 50 fb\textsuperscript{-1} integrated luminosity, respectively.  Tables \ref{tab:fccmu1500} and \ref{tab:fccmu3000} give detailed compositeness scale limit results with their sensitivities for both  $\mu p$ colliders.

Regarding the last collider option in Table \ref{tab:fccMup} with 63.2 TeV center-of-mass energy, figure \ref{fig:ci_mup20000} depicts the same pattern as other collider options that the lowest compositeness scale limits are revealed for $\Lambda^-_{RL}$, and the highest compositeness scale limits appeared for $\Lambda^+_{LL}$ at all confidence levels. The final integrated luminosity  value is 100 fb\textsuperscript{-1} for the 	$\mu$20000$\otimes$FCC collider at the end of the ten years. The highest discovery, observation, and exclusion limits are obtained as 225.7 $\pm$ 1.9\% TeV,  269.0 $\pm$ 2.0\%  TeV, and 311.3 $\pm$ 2.1\% TeV with constructive  $\Lambda_{LL}$, and  the lowest limits of confidence levels are achieved as 139.1 $\pm$ 1.3\% TeV,  155.7 $\pm$ 1.3\% TeV, and  170.3 $\pm$ 1.3\%  TeV with destructive $\Lambda_{RL}$, respectively.  Table \ref{tab:fccmu20000} lists detailed compositeness scale limits with sensitivity for all confidence levels at $\mu$20000$\otimes$FCC collider. 

\onecolumngrid

\begin{table}[h!]
	\caption{Attainable limits on the compositeness scales with sensitivity at the FCC-based $\mu p$ collider with $\sqrt{s}$ = 17.3 TeV for the first, the  fifth, and the tenth years' luminosities. }
	\resizebox{\textwidth}{!}{\begin{tabular}{cccccccccc}\hline
			\multicolumn{2}{l}{\textbf{$\mu$1500$\otimes$FCC}} &
			\multicolumn{8}{c}{$\Lambda \pm \delta \%$ (TeV)} \\ \hline
			\begin{tabular}[c]{@{}c@{}}$\mathcal{L}_{int}$ \\ (fb\textsuperscript{-1})\end{tabular} &
			CL &
			$\Lambda^{+}_{LL}$ &
			$\Lambda^{-}_{LL}$ &
			$\Lambda^{+}_{RR}$ &
			$\Lambda^{-}_{RR}$ &
			$\Lambda^{+}_{LR}$ &
			$\Lambda^{-}_{LR}$ &
			$\Lambda^{+}_{RL}$ &
			$\Lambda^{-}_{RL}$ \\ \hline
			\multirow{3}{*}{5} &
			5$\sigma$ &
			52.7 ± 4.0\% &
			41.3 ± 2.7\% &
			51.4 ± 3.9\% &
			42.0 ± 2.9\% &
			38.4 ± 3.6\% &
			35.8 ± 3.1\% &
			38.9 ± 3.5\% &
			35.6 ± 3.0\% \\
			&
			3$\sigma$ &
			62.4 ± 4.2\% &
			45.8 ± 2.4\% &
			60.5 ± 4.0\% &
			46.9 ± 2.6\% &
			44.4 ± 3.6\% &
			40.4 ± 2.9\% &
			44.9 ± 3.5\% &
			40.0 ± 2.8\% \\
			&
			2$\sigma$ &
			72.0 ± 4.4\% &
			49.4 ± 2.3\% &
			69.3 ± 4.2\% &
			50.8 ± 2.4\% &
			50.1 ± 3.7\% &
			44.2 ± 2.7\% &
			50.4 ± 3.6\% &
			43.7 ± 2.7\% \\ \hline
			\multirow{3}{*}{25} &
			5$\sigma$ &
			69.2 ± 2.9\% &
			48.4 ± 1.5\% &
			66.7 ± 2.8\% &
			49.8 ± 1.6\% &
			48.4 ± 2.4\% &
			43.2 ± 1.8\% &
			48.8 ± 2.3\% &
			42.7 ± 1.8\% \\
			&
			3$\sigma$ &
			83.4 ± 3.1\% &
			52.9 ± 1.4\% &
			79.8 ± 2.9\% &
			54.8 ± 1.5\% &
			56.5 ± 2.5\% &
			48.2 ± 1.7\% &
			56.7 ± 2.4\% &
			47.5 ± 1.7\% \\
			&
			2$\sigma$ &
			97.6 ± 3.2\% &
			56.5 ± 1.3\% &
			92.7 ± 3.1\% &
			58.8 ± 1.4\% &
			64.1 ± 2.6\% &
			52.3 ± 1.7\% &
			64.1 ± 2.5\% &
			51.6 ± 1.7\% \\ \hline
			\multirow{3}{*}{50} &
			5$\sigma$ &
			78.4 ± 2.5\% &
			51.5 ± 1.2\% &
			75.2 ± 2.4\% &
			53.1 ± 1.3\% &
			53.7 ± 2.1\% &
			46.5 ± 1.5\% &
			54.0 ± 2.0\% &
			45.9 ± 1.4\% \\
			&
			3$\sigma$ &
			95.3 ± 2.7\% &
			56.0 ± 1.1\% &
			90.7 ± 2.6\% &
			58.2 ± 1.2\% &
			62.9 ± 2.2\% &
			51.7 ± 1.4\% &
			63.0 ± 2.1\% &
			51.0 ± 1.4\% \\
			&
			2$\sigma$ &
			112.3 ± 2.8\% &
			59.6 ± 1.0\% &
			106.1 ± 2.7\% &
			62.4 ± 1.1\% &
			71.8 ± 2.3\% &
			56.1 ± 1.4\% &
			71.7 ± 2.2\% &
			55.4 ± 1.4\% \\ \hline
		\end{tabular}\label{tab:fccmu1500} }
\end{table}



\begin{table}[h!]
	\caption{Attainable limits on the compositeness scales with sensitivity at the FCC-based $\mu p$ collider with $\sqrt{s}$ = 24.5 TeV for the first, the  fifth, and the tenth years' luminosities.}
	\resizebox{\textwidth}{!}{\begin{tabular}{cccccccccc} \hline
			\multicolumn{2}{l}{\textbf{$\mu$3000$\otimes$FCC}} &
			\multicolumn{8}{c}{$\Lambda \pm \delta \%$ (TeV)} \\ \hline
			\begin{tabular}[c]{@{}c@{}}$\mathcal{L}_{int}$ \\ (fb\textsuperscript{-1})\end{tabular} &
			CL &
			$\Lambda^{+}_{LL}$ &
			$\Lambda^{-}_{LL}$ &
			$\Lambda^{+}_{RR}$ &
			$\Lambda^{-}_{RR}$ &
			$\Lambda^{+}_{LR}$ &
			$\Lambda^{-}_{LR}$ &
			$\Lambda^{+}_{RL}$ &
			$\Lambda^{-}_{RL}$ \\ \hline
			\multirow{3}{*}{5} &
			5$\sigma$ &
			71.4 ± 3.9\% &
			56.8 ± 2.6\% &
			70.0 ± 3.8\% &
			58.5 ± 2.7\% &
			50.9 ± 3.5\% &
			47.8 ± 2.8\% &
			51.1 ± 3.4\% &
			47.5 ± 2.9\% \\
			&
			3$\sigma$ &
			84.2 ± 4.1\% &
			62.9 ± 2.4\% &
			82.0 ± 3.9\% &
			65.0 ± 2.5\% &
			58.8 ± 3.5\% &
			53.5 ± 2.8\% &
			58.9 ± 3.5\% &
			53.3 ± 2.8\% \\
			&
			2$\sigma$ &
			96.5 ± 4.2\% &
			67.8 ± 2.3\% &
			93.5 ± 4.1\% &
			70.1 ± 2.3\% &
			66.1 ± 3.6\% &
			58.5 ± 2.7\% &
			66.0 ± 3.5\% &
			58.3 ± 2.7\% \\ \hline
			\multirow{3}{*}{25} &
			5$\sigma$ &
			92.8 ± 2.8\% &
			66.4 ± 1.5\% &
			90.2 ± 2.7\% &
			68.7 ± 1.5\% &
			64.0 ± 2.4\% &
			57.1 ± 1.8\% &
			64.0 ± 2.3\% &
			56.9 ± 1.8\% \\
			&
			3$\sigma$ &
			111.0 ± 2.9\% &
			72.7 ± 1.4\% &
			107.1 ± 2.8\% &
			75.4 ± 1.5\% &
			74.5 ± 2.5\% &
			63.8 ± 1.8\% &
			74.1 ± 2.4\% &
			63.6 ± 1.8\% \\
			&
			2$\sigma$ &
			129.0 ± 3.1\% &
			77.7 ± 1.3\% &
			123.6 ± 3.0\% &
			80.8 ± 1.4\% &
			84.6 ± 2.6\% &
			69.8 ± 1.8\% &
			83.7 ± 2.5\% &
			69.5 ± 1.8\% \\ \hline
			\multirow{3}{*}{50} &
			5$\sigma$ &
			104.7 ± 2.4\% &
			70.7 ± 1.2\% &
			101.2 ± 2.3\% &
			73.2 ± 1.2\% &
			70.9 ± 2.1\% &
			61.6 ± 1.5\% &
			70.6 ± 2.0\% &
			61.4 ±  1.5\% \\
			&
			3$\sigma$ &
			126.1 ± 2.6\% &
			77.0 ± 1.1\% &
			121.0 ± 2.5\% &
			80.0 ± 1.2\% &
			83.0 ± 2.2\% &
			68.9 ± 1.5\% &
			82.2 ± 2.1\% &
			68.6 ± 1.5\% \\
			&
			2$\sigma$ &
			147.5 ± 2.7\% &
			82.2 ± 1.1\% &
			140.6 ± 2.6\% &
			85.6 ± 1.1\% &
			94.8 ± 2.3\% &
			75.5 ± 1.6\% &
			93.2 ± 2.2\% &
			75.2 ± 1.5\%\\ \hline
		\end{tabular}\label{tab:fccmu3000}}
\end{table}


\twocolumngrid

\onecolumngrid

\begin{table}[h!]
	\caption{Attainable limits on the compositeness scales with sensitivity at the  FCC-based $\mu p$ collider with $\sqrt{s}$ = 63.2 TeV for the first, the  fifth, and the tenth years' luminosities.}
	\resizebox{\textwidth}{!}{\begin{tabular}{cccccccccc} \hline
			\multicolumn{2}{l}{\textbf{$\mu$20000$\otimes$FCC}} &
			\multicolumn{8}{c}{$\Lambda \pm \delta \%$ (TeV)} \\ \hline
			\begin{tabular}[c]{@{}c@{}}$\mathcal{L}_{int}$ \\ (fb\textsuperscript{-1})\end{tabular} &
			CL &
			$\Lambda^{+}_{LL}$ &
			$\Lambda^{-}_{LL}$ &
			$\Lambda^{+}_{RR}$ &
			$\Lambda^{-}_{RR}$ &
			$\Lambda^{+}_{LR}$ &
			$\Lambda^{-}_{LR}$ &
			$\Lambda^{+}_{RL}$ &
			$\Lambda^{-}_{RL}$ \\ \hline
			\multirow{3}{*}{10} &
			5$\sigma$ &
			157.0 ± 2.9\% &
			130.3 ± 2.4\% &
			153.8 ± 3.0\% &
			132.0 ± 2.4\% &
			113.0 ± 2.9\% &
			106.4 ± 2.6\% &
			113.1 ± 2.9\% &
			106.8 ± 2.5\% \\
			&
			3$\sigma$ &
			183.6 ± 3.2\% &
			144.8 ± 2.1\% &
			179.2 ± 3.2\% &
			147.7 ± 2.2\% &
			130.6 ± 2.9\% &
			120.4 ± 2.5\% &
			130.2 ± 2.8\% &
			120.5 ± 2.4\% \\
			&
			2$\sigma$ &
			208.9 ± 3.3\% &
			156.8 ± 2.0\% &
			203.9 ± 3.4\% &
			160.5 ± 2.1\% &
			146.7 ± 3.0\% &
			132.2 ± 2.4\% &
			145.5 ± 2.9\% &
			132.1 ± 2.3\% \\ \hline
			\multirow{3}{*}{50} &
			5$\sigma$ &
			201.5 ± 2.2\% &
			153.5 ± 1.3\% &
			196.6 ± 2.2\% &
			157.0 ± 1.4\% &
			142.1 ± 2.0\% &
			128.9 ± 1.6\% &
			141.1 ± 1.9\% &
			128.8 ± 1.6\% \\
			&
			3$\sigma$ &
			238.5 ± 2.3\% &
			168.9 ± 1.2\% &
			233.3 ± 2.4\% &
			173.4 ± 1.3\% &
			164.9 ± 2.0\% &
			144.6 ± 1.5\% &
			162.6 ± 1.9\% &
			144.3 ± 1.5\% \\
			&
			2$\sigma$ &
			274.7 ± 2.4\% &
			181.1 ± 1.1\% &
			269.6 ± 2.5\% &
			187.0 ± 1.3\% &
			186.5 ± 2.1\% &
			158.0 ± 1.5\% &
			182.5 ± 2.0\% &
			157.7 ± 1.5\% \\ \hline
			\multirow{3}{*}{100} &
			5$\sigma$ &
			225.7 ± 1.9\% &
			163.9 ± 1.1\% &
			220.5 ± 1.9\% &
			168.1 ± 1.1\% &
			157.1 ± 1.7\% &
			139.4 ± 1.3\% &
			155.3 ± 1.6\% &
			139.1 ± 1.3\% \\
			&
			3$\sigma$ &
			269.0 ± 2.0\% &
			179.3 ± 1.0\% &
			263.8 ± 2.1\% &
			185.0 ± 1.1\% &
			183.1 ± 1.8\% &
			156.0 ± 1.3\% &
			179.4 ± 1.6\% &
			155.7 ± 1.3\% \\
			&
			2$\sigma$ &
			311.3 ± 2.1\% &
			191.3 ± 0.9\% &
			307.2 ± 2.2\% &
			199.0 ± 1.0\% &
			208.0 ± 1.8\% &
			170.4 ± 1.3\% &
			202.0 ± 1.7\% &
			170.3 ± 1.3\%\\ \hline
		\end{tabular}\label{tab:fccmu20000} }
\end{table}

\twocolumngrid



\section{\label{sec:conclusion} Conclusion}
We carried out a contact interaction study at FCC-based muon-proton colliders. In our calculations, we considered four center-of-mass energies $\mu p$ collider options. FCC-based $\mu p$ collider that the highest center-of-mass energy (63.2 TeV) option with the luminosity 100 fb\textsuperscript{-1} reveals the largest attainable compositeness scales. This machine will allow discovery up to  225.7 $\pm$ 1.9\% TeV, observation 269.0 $\pm$ 2.0\% TeV, and exclusion 311.3 $\pm$ 2.1 \%TeV for $\Lambda^+_{LL}$ (constructive interference) in four-fermion contact interactions.  When we consider destructive interference for left-left helicity at this collider, discovery, observation, and exclusion limits are calculated as 163.9 $\pm$ 1.1\%  TeV, 179.3 $\pm$ 1.0\%  TeV, and 191.3 $\pm$ 0.9\% TeV, respectively. As can be seen from Reference \cite{Apanasevich:2013cta}, it has been demonstrated that in future proton-proton colliders, with an integrated luminosity value of 3000 fb\textsuperscript{-1}, an exclusion limit of approximately 45 TeV can be imposed on the compositeness scale (LL, RR) by contact interactions in the 33 TeV center-of-mass energy option. With the same integrated luminosity value, for the option with a center-of-mass energy of 100 TeV, an exclusion limit of 125 TeV was envisaged to be put on the compositeness scale. As shown in Table \ref{tab:fccmu3000}, the limit imposed on the compositeness scale is higher in the muon-proton collider in our study with a 24.5 TeV center of mass-energy and 50 fb\textsuperscript{-1} integrated luminosity. Moreover, the muon-proton collider option with 100 fb\textsuperscript{-1} integrated luminosity with 63.5 TeV center-of-mass energy can also introduce a higher exclusion limit to the compositeness scale than the 100 TeV center-of-mass collider in reference \cite{Apanasevich:2013cta}. In addition, the compositeness scale could be examined up to 100 TeV for contact interactions in the FCC-eh collider \cite{fcc2019fcc:hh}. We indicated that FCC-$\mu p$ colliders could push the limits of the compositeness scale in contact interactions higher than references \cite{fcc2019fcc:hh}, and \cite{Apanasevich:2013cta}. Our findings show that FCC-based muon-proton colliders have a great potential regarding four-fermion contact interactions than LHC, ILC ,CLIC, FCC-eh, and future $pp$ colliders.  As a result, FCC-based $\mu p$ colliders will be an exceptional alternative for researching four-fermion contact interactions.

\begin{acknowledgments}
The authors are obliged to the Usak University Energy, Environment, and Sustainability Application and Research Center for support. Figures in this paper are created by using ROOT software \cite{brun1997}. 
\end{acknowledgments}

\newpage
\bibliography{contactIntacts.bib}

\end{document}